\documentclass[preprint2]{aastex62}

\usepackage{multirow}
\usepackage{comment}

\newcommand{\obj}{IRAS F05189-2524}
\newcommand{\xmm}{\textit{XMM-Newton}}
\newcommand{\nus}{\textit{NuSTAR}}
\newcommand{\e}[1]{\ensuremath{\times10^{#1}}}

\accepted{October 16, 2019}
\submitjournal{ApJ}

\shorttitle{X-ray Wind Driving Galactic Outflow in \obj}
\shortauthors{Smith et al.}

\begin{document}

\title{Discovery of an X-ray Quasar Wind Driving the Cold Gas Outflow in the Ultraluminous Infrared Galaxy \obj}

\author[0000-0001-5626-5209]{Robyn N. Smith}
\affil{Department of Astronomy, University of Maryland, College Park, MD 20742, USA}

\author[0000-0002-6562-8654]{Francesco Tombesi}
\affil{Department of Physics, University of Rome `Tor Vergata', Via della Ricerca Scientifica 1, I-00133 Rome, Italy}
\affil{Department of Astronomy, University of Maryland, College Park, MD 20742, USA}
\affil{NASA/Goddard Space Flight Center, Code 662, Greenbelt, MD 20771, USA }
\affil{INAF Osservatorio Astronomico di Roma, Via Frascati 33, 00078 Monteporzio Catone, Italy}

\author[0000-0002-3158-6820]{Sylvain Veilleux}
\affil{Department of Astronomy, University of Maryland, College Park, MD 20742, USA}
\affil{Joint Space-Science Institute, University of Maryland, College Park, MD 20742, USA}
\affil{Institute of Astronomy and Kavli Institute for Cosmology Cambridge, University of Cambridge, Cambridge, CB3 0HA, United Kingdom}

\author{Anne M. Lohfink}
\affil{Montana State University, Department of Physics, P.O. Box 173840, Bozeman, MT 59717, USA}

\author{Alfredo Luminari}
\affil{Department of Physics, University of Rome `Tor Vergata', Via della Ricerca Scientifica 1, I-00133 Rome, Italy}
\affil{INAF Osservatorio Astronomico di Roma, Via Frascati 33, 00078 Monteporzio Catone, Italy}

\correspondingauthor{Robyn N. Smith}
\email{rnsmith@astro.umd.edu}

\begin{abstract}
We present new \xmm\ and \nus\ observations of the galaxy merger \obj\ which is classified as an ultra-luminous infrared galaxy (ULIRG) and optical Seyfert 2 at $z$ = 0.0426. We test a variety of spectral models which yields a best-fit consisting of an absorbed power law with emission and absorption features in the Fe K band. Remarkably, we find evidence for a blueshifted Fe K absorption feature at $E$ = 7.8 keV (rest-frame) which implies an ultra-fast outflow (UFO) with $v_{\mathrm{out}} = 0.11\ \pm\ 0.01c$. We calculate that the UFO in \obj\ has a mass outflow rate of $\dot{M}_{\mathrm{out}}\ \gtrsim 1.0\ M_\sun$ yr$^{-1}$, a kinetic power of $\dot{E}_{\mathrm{K}} \gtrsim$ 8\% $L_{\mathrm{AGN}}$, and a momentum rate (or force) of $\dot{P}_{\mathrm{out}}\ \gtrsim 1.4\  L_{\mathrm{AGN}}/c$. Comparing the energetics of the UFO to the observed multi-phase outflows at kiloparsec scales yields an efficiency factor of $f\sim0.05$ for an energy-driven outflow. Given the uncertainties, however, we cannot exclude the possibility of a momentum-driven outflow. Comparing \obj\ with nine other objects with observed UFOs and large-scale galactic outflows suggests that there is a range of efficiency factors for the coupling of the energetics of the nuclear and galaxy-scale outflows that likely depend on specific physical conditions in each object. 
\end{abstract}


\section{Introduction} \label{sec:intro}

In one possible evolutionary scenario, gas-rich galaxies merge together to form an obscured ultraluminous infrared galaxy (ULIRG) which evolves into a dusty quasar and then eventually an exposed optical quasar after shedding its gas and dust cocoon (e.g., \citealt{sanders88, veilleux02, veilleux09a, veilleux09b, hopkins06}). This scenario may account for the intimate link between the mass of the stellar spheroid component of the host galaxy and that of the central supermassive black hole (SMBH) (e.g., \citealt{silk98, ferrarese00, gebhardt00}) by invoking negative feedback of the active galactic nucleus (AGN) i.e. the AGN drives galactic winds which in turn may be able to quench the growth of both the SMBH and the stellar component of the host (e.g. \citealt{fabian99, king03, kingpounds03}). 

Star formation is inhibited if the cold molecular gas out of which stars form is affected by such outflows. Far-infrared molecular spectroscopy of ULIRGs has revealed highly blueshifted absorption features indicative of high-velocity molecular outflows on scales of hundreds of parsecs which imply significant mass outflow rates \citep{sturm11,veilleux13, veilleux17, gonzalf17, rupke17}. Most models explaining the origin of these galactic outflows require a very fast ($v_{\mathrm{out}} \sim 0.1c$) initial AGN accretion disk wind which shocks the surrounding interstellar medium (ISM) and forms a hot bubble which moves the molecular material (see \citealt{kingpounds15} and references therein). The shock-driven galactic outflow can be divided into two distinct regimes: momentum-driven and energy-driven. 

Momentum-driven outflows occur when the kinetic energy of the wind is mostly radiated away, in which case, only ram pressure exerts work on the surrounding ISM. Energy-driven outflows occur if the shocked ISM is not efficiently cooled and expands adiabatically as a hot bubble. The momentum rate of an energy-driven outflow is expected to be larger than that of a momentum-driven outflow and may approach values of $\dot{P} \simeq 10\ L_{\mathrm{AGN}}/c$ which is consistent with observations of several ULIRGs \citep{sturm11, cicone14, gonzalf17}. 

Galactic-scale outflows are common in U/LIRGs and often involve several gas phases: the molecular gas (e.g., 
\citealt{veilleux13, gonzalf17, fluetsch18}), the neutral atomic gas \citep{rupke13, rupke17, teng13}, the warm ionized gas \citep{rupke13, rupke17}, and sometimes even the hot ionized gas (\citealt{nardini13, veilleux14, paggi17, liu19}). Conversely, outflows inferred from blueshifted \ion{Fe}{25}/\ion{}{26} absorption lines in the X-ray band at rest-frame energies $E >$ 7 keV are observed in AGN at sub-parsec scales consistent with an accretion disk interpretation. These ultra-fast outflows (UFOs; \citealt{tombesi10, tombesi11, tombesi14, tombesi15, gofford13, longinotti15, nardini15, parker17}), have outflow velocities which are mildly relativistic ($v_{\mathrm{out}} \sim 0.1c$). Confirming both a large-scale galactic outflow and sub-parsec scale accretion disk wind in the same object presents observational challenges requiring simultaneous detection of the outflow in the X-rays and at lower energies (mm-optical-IR). 

IRAS F11119+3257 was the first such source in which both outflows were confirmed. Galactic outflows were found using OH absorption measurements with \textit{Herschel} \citep{veilleux13} and confirmed with CO(1--0) emission line measurements from deep \textit{ALMA} observations \citep{veilleux17}. The UFO was initially detected with \textit{Suzaku} \citep{tombesi15} and later confirmed with \textit{NuSTAR} observations \citep{tombesi17}. Mrk 231 is the second known object whose outflows were confirmed using \textit{IRAM}, \textit{Chandra}, and \nus\  \citep{feruglio15}. Therefore, it is imperative to extend such studies to other sources in order to quantify the occurrence of such phenomena.

\section{\obj} \label{sec:obj}

\obj\ is a well-studied, nearby ($z=0.0426$), late-stage merger ULIRG \citep{veilleux02, veilleux06}. It is an optical Seyfert 2 \citep{veilleux99a}, but contains hidden broad-line Pa$\beta$ in the near-infrared \citep{veilleux99b}. With $\sim$70\% of its bolometric luminosity ($L_{\mathrm{bol}} \sim 10^{12}\ L_{\odot}$) attributed to its AGN \citep{veilleux09a}, the AGN in \obj\ is considered a quasar. A high-velocity, large-scale outflow has been detected in the neutral, ionized, and molecular gas phases \citep{rupke05a, westmoquette12, bellocchi13, teng13, veilleux13, gonzalf17, rupke17}.

In the X-ray, \obj\ is one of the brightest local ULIRGs. Archival \xmm\ and \textit{Chandra} observations derive an $E=$ 2--10 keV continuum luminosity of $\sim10^{43}$ erg s$^{-1}$ \citep{teng09}. The X-ray flux of \obj\ is known to vary. The $E=$ 0.5--2 keV flux was relatively constant during \xmm\ observations in 2001 March, \textit{Chandra} observations in 2001 October and 2002 January, and \textit{Suzaku} observations in 2006 April. The $E=$ 2--10 keV flux, however, was a factor of $\sim$30 lower in the 2006 \textit{Suzaku} than previously measured in the \xmm\ and \textit{Chandra} observations in 2001-02. In addition to the drop in flux, the 2006 \textit{Suzaku} observation revealed a prominent $E=$ 6.4 keV Fe K emission line not seen in the 2001-02 observations \citep{teng09}. Observations by \textit{ASCA} in 1995 and \textit{BeppoSAX} in 1999 found statistically significant unresolved iron line emission, but also confirmed strong continuum variability above $E=$ 2 keV between the two observations \citep{severgnini01}. \obj\ was observed by \nus\ in 2013 February (21 ks) and October (25 and 8 ks) with a coordinated \xmm\ observation during the 2013 October observation (31 ks; \citealt{teng15}). Minor flux variations detected between these observations were not found to be statistically significant, and the $E=$ 2--10 keV flux was again consistent with the ``high'' state of the 2001-02 observations \citep{teng15}. 

\obj\ was detected by \textit{Swift} BAT with a significance of 6 $\sigma$ at $E=$ 14--195 keV and 4.2 $\sigma$ at $E=$ 24--35 keV \citep{koss13}. In re-analyzing the 2013 \nus\ and \xmm\ observations, \citet{xu17} find that \obj\ may be modeled above $E=$ 2 keV by a broad iron line disk reflection. \citet{xu17} also find that possible features indicative of a high-velocity outflow in the Fe K band are not statistically required after the fit with a relativistic reflection dominated spectral model. Data of higher quality are needed to confirm the possible existence of these spectral features.

\section{Observations and Data Reduction} \label{sec:data}

\subsection{\xmm}\label{sec:xmm}

\obj\ was observed by \xmm\ for 98 ks on 2016 Sept. 6--7 (ObsID 0790580101). The observations were reduced using standard procedures with the \xmm\ Science Analysis System v16.1.0. Soft proton flares were removed, and only single and double events were retained for the pn while single through quadruple events were retained for the MOS. The source was extracted using a 40'' radius circular region. The background was estimated from a source-free sky region of the same size. For the pn background, special care was taken to ensure that the background region was not located on parts of the CCD where there are known instrumental X-ray fluorescent lines \citep{freyberg04}, particularly the Cu-K$\alpha$ line around 8 keV. The final good exposure time for the pn was 74.3 ks. The MOS1 and MOS2 observations were reduced separately. Each MOS spectrum and light curve was inspected individually, and finding no gross variability between the two, they were combined using \texttt{epicspeccombine}. The final good exposure time for the combined MOS spectrum is 94.7 ks. Table \ref{tab:obsinfo} provides the final good exposure times and count rates for the \xmm\ observation. The final spectrum for both the pn and MOS were grouped to a minimum of 50 counts per bin in order to ensure the use of the $\chi^2$ statistics.

\subsection{\nus}\label{sec:nus}
\obj\ was observed by \nus\ for 144 ks on 2016 Sept. 5--8 (ObsID 60201022002). Spectra were created using HEAsoft version 6.22 and CALDB version `20171002' after initially producing cleaned event files with the tool \texttt{nupipeline}. For the screening parameters, we assumed ``saacalc=2 saamode=optimized tentacle=yes" based on the \nus\ SAA filtering report. From the cleaned event files, spectra and corresponding response matrices were then created using the \texttt{nuproducts} tool. The source region was chosen to be circular with a 60" radius, the background region was also circular with 121" radius. The resulting spectra have a net exposure of 144.1 ks for focal plane module (FPM) A and 143.9 ks for the FPMB. Due to differing orbits, the \nus\ observation is only strictly concurrent with \xmm\ for 45 ks. Table \ref{tab:obsinfo} provides the final good exposure times and count rates for the \nus\ observation. All FPMA and FPMB spectra were grouped to a minimum of 25 counts per bin in order to ensure he use of the $\chi^2$ statistics.

\begin{deluxetable*}{lrrr}[]
\tablecaption{Exposure times and count rates for \xmm\ observation (ID 0790580101) and \nus\ observation (ID 60201022002) of \obj.\label{tab:obsinfo}}
\tablecolumns{4}
\tablewidth{\textwidth}
\tablehead{
\colhead{Instrument} &
\colhead{Exposure}  & 
\colhead{Count Rate} &
\colhead{Count Rate}\\
& \colhead{(ks)} & \colhead{0.5-2 keV} & \colhead{2-10 keV$^a$} 
}
\startdata
EPIC pn  & 74.33 & 0.393 & 0.315\\
EPIC MOS & 94.71 & 0.251  & 0.212\\
FPMA (full) & 144.1 & & 0.067 \\
FPMB (full) & 143.9 & & 0.061\\
FPMA (simultaneous) & 45.39 & & 0.073\\
FPMB (simultaneous) & 45.31 & & 0.066\\
\enddata
\tablenotetext{a}{Count rates for \nus\ are calculated between 3--10 keV.}
\end{deluxetable*}

\section{\xmm\ Spectral Analysis} \label{sec:spectra}

\begin{figure}
\epsscale{1.35}
\plotone{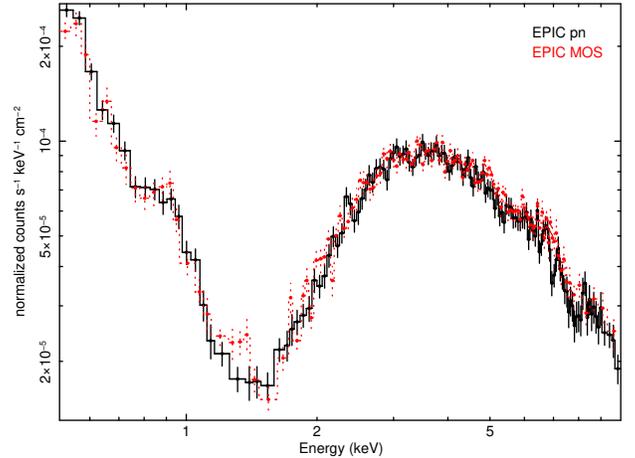}
\caption{The \xmm\ EPIC spectrum of \obj\ from 0.5--10 keV in the rest frame ($z = 0.0426$). Additional binning has been applied for visual purposes.}
\label{fig:data}
\end{figure}

We perform our spectral analysis using XSPEC v12.10c \citep{arnaud96} using $\chi^2$ statistics. All models take into account Galactic absorption with the \texttt{tbabs} model \citep{wilms00} using a Galactic column density of $N_{\mathrm{H,Gal}} = 1.66\times10^{20}$ cm$^{-2}$ \citep{kalberla05}. All parameters are given in the rest frame of \obj\ ($z$ = 0.0426). The full \xmm\ EPIC spectrum of \obj\ from 0.5--10 keV is presented in Fig. \ref{fig:data}. All errors and limits are given at a level of 90\% ($\Delta\chi^2 = 2.7$ for one degree of freedom). Statistical calculations were performed using XSPEC \texttt{error} and \texttt{steppar} commands avoiding local minima when searching $\chi^2$ space. The difference in sensitivity of the pn and MOS spectra are due to the difference in effective area. The effective area of the MOS decreases more rapidly at higher energies than the pn.

\subsection{Broad-band Modeling} \label{sec:broad}
We begin by joint modeling the EPIC pn and MOS spectra from 0.5--10 keV with a simple power law. This provides a poor fit ($\chi^2_{\mathrm{red}}$ = $\chi^2$/$\nu$ = 11.93) and is not considered further. The spectrum is indicative of a soft X-ray absorber (see Fig. \ref{fig:data}), so our next model invokes a full covering neutral absorber (\texttt{zwabs} in XSPEC). While this provides a better fit ($\chi^2_{\mathrm{red}}$ = 6.47), it is clear that the model is not accounting for any emission that is present at soft X-ray energies ($E < 2$ keV) and is not considered further. 

\begin{figure*}
\plotone{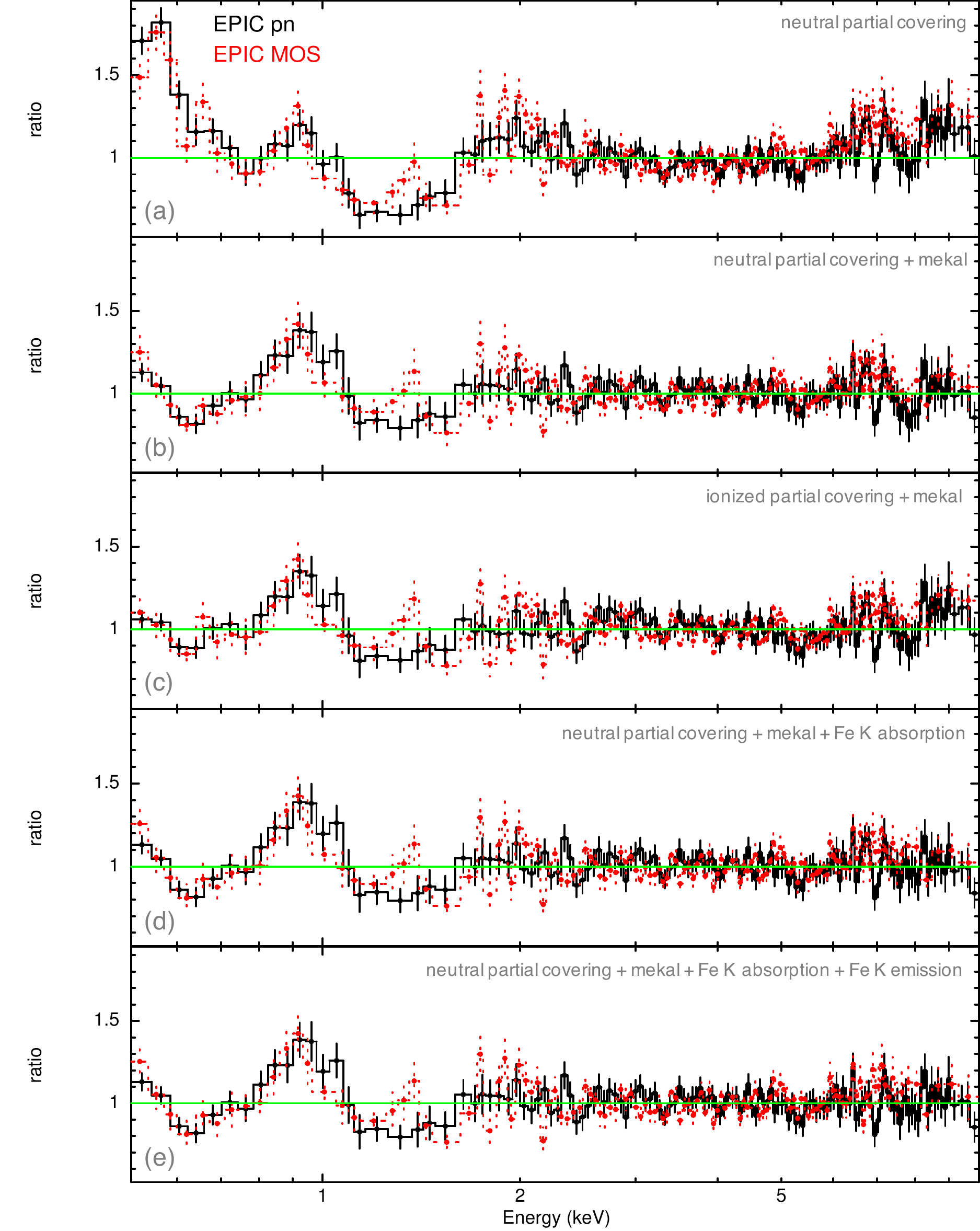}
\caption{Data-to-model ratios for broad-band models; EPIC pn is plotted in black, the MOS in red. (a) a neutral partial covering absorber by itself does not account for excess emission below $E$=1 keV and underestimates the continuum above $E$=8 keV; (b) adding a \texttt{mekal} component to account for hot diffuse gas improves issues seen in (a); (c) testing a ionized neutral absorber results in an underestimate of the continuum above $E$=8 keV; (d) a Gaussian is added to (b) to model an absorption feature at $E$=7.8 keV; (e) a Gaussian is added to (d) to model an emission feature at $E$=6.7 keV.}
\label{fig:broad}
\end{figure*}

We then consider a neutral partial covering absorber (\texttt{zpcfabs}) which provides a significant improvement in the overall fit ($\chi^2_{\mathrm{red}}$ = 1501/948 = 1.58; see Fig. \ref{fig:broad}a) although there is still excess emission at soft energies. This model has a column density $N_{\mathrm{H}} = (8.54\ \pm\ 0.12$) \e{22} cm$^{-2}$, and the photon index, while high ($\Gamma$ = 2.29 $\pm\ 0.01$), is not unreasonable given the large range of previously published values for \obj\ \citep{risaliti00, ptak13, teng15, xu17}. We also test a continuum scattering model using two power laws with the same photon index, one with full neutral absorption and one with no absorption. The fit of this model is comparable to that with neutral partial covering absorption with no clear preference for either model. Although these models are phenomenologically distinct, they are mathematically equivalent, and we will continue with our spectral analysis using neutral partial covering absorption.

Since \obj\ is a ULIRG, we add a \texttt{mekal} component to account for the hot diffuse gas likely present in the host galaxy (see Fig. \ref{fig:broad}b). The \texttt{mekal} component has a plasma temperature of $kT = 0.181\ \pm\ 0.004$ keV while the neutral partial covering absorber has a column density of $N_{\mathrm{H}} = (7.29\ \pm\ 0.10$) \e{22} cm$^{-2}$. This improves the fit to $\chi^2_{\mathrm{red}}$ = 1050/946 = 1.11 in addition to yielding a photon index of $\Gamma$ = 1.97 $\pm\ 0.01$, much closer to the canonical value of $\Gamma$ = 2 (e.g., \citealt{nandrapounds94, reevesturner00}). Both neutral partial covering absorber models (with and without the \texttt{mekal} component) have a covering fraction of 98\% with full covering excluded at the 90\% level.

Next, we test an ionized partial covering absorber (\texttt{zxipcf}), which slightly improves the fit ($\chi^2_{\mathrm{red}}$ = 1034/945 = 1.09; see Fig. \ref{fig:broad}c). The column density increases to $N_{\mathrm{H}} = (11.06\ \pm\ 1.10$) \e{22} cm$^{-2}$ with a low ionization parameter of log $\xi$ = 0.59 $\pm\ 0.05$ erg s$^{-1}$ cm and a covering fraction of 99\% (full covering remains excluded at the 90\% level). The plasma temperature of the \texttt{mekal} component decreases slightly to $kT$ = 0.147 $^{+0.016}_{-0.024}$ keV. Now, however, the continuum above $E =$ 8 keV is noticeably underestimated (see Fig. \ref{fig:broad}c) while the photon index has steepened ($\Gamma$ = 2.49 $\pm\ 0.12$). This is consistent with the ionized partial covering absorber compromising the continuum fit for the sake of the large contribution to the residuals at lower energies. For these reasons, we discard the model with the ionized partial covering absorber. 

From the data-to-model ratios in Fig. \ref{fig:broad}, we find evidence for a possible absorption feature between $E\sim7-8$ keV. We approximate this feature by adding a Gaussian to our model with a neutral partial covering absorber and \texttt{mekal} component (see Fig. \ref{fig:broad}d). The center of the line is located at $E = 7.81\ \pm\ 0.10$ keV with a width of $\sigma$ = 103 eV (90\% upper limit $\sigma <$ 248 eV) and an equivalent width of 56 $^{+37}_{-34}$ eV. This improves the fit by $\Delta\chi^2/\Delta\nu$ = 9/3 which corresponds to a statistical requirement of 97\% according to the F-test ($>2\sigma$). The column density is $N_{\mathrm{H}} = (7.22\ \pm\ 0.10$) \e{22} cm$^{-2}$, covering fraction is 98\%, photon index $\Gamma$ = 1.94 $\pm\ 0.01$, and plasma temperature $kT = 0.181\ \pm\ 0.004$ keV. 

In our last broad-band model, we add an unresolved Gaussian emission line at $E = 6.70\ \pm\ 0.06$ keV with a width frozen to $\sigma$ = 10 eV, consistent with iron K lines of highly ionized iron (\ion{Fe}{18} and above; \citealt{kallman04}; see Fig. \ref{fig:broad}e). The addition of the emission line narrows the absorption feature to $\sigma$ = 78 eV (90\% upper limit $\sigma <$ 240 eV). The equivalent width of the absorption feature also decreases to 46 $^{+36}_{-39}$ eV while the equivalent width of the emission feature is 35 $\pm17$ eV. Using an F-test, the addition of a second Gaussian is statistically significant at a level of 99.7\% ($\sim3\sigma$). Other model parameters remain largely unchanged by the inclusion of an emission feature. The parameters of this best-fitting broad-band model ($\chi^2_{\mathrm{red}}$ = 1028/941 = 1.09) are presented in Table \ref{tab:mod3c}. 

\begin{deluxetable*}{llcccc}[t!]
\tablecaption{Parameters for the best-fitting broad-band model. All errors are given at the 90\% level while limits are given at 90\%. \label{tab:mod3c}}
\tablecolumns{6}
\tablewidth{\textwidth}
\tablehead{
\colhead{Component} &
\colhead{Parameter} &
\colhead{Unit} &
\colhead{Model}  & 
\colhead{$\chi^2$/$\nu$} & 
\colhead{$\Delta\chi^2$/$\Delta\nu$} 
}
\startdata
zpowerlw & $\Gamma$ & & 1.97 $\pm$ 0.01 & 11334/950 & \ldots \\
                & $z$\tablenotemark{a} & & 0.0426 & & \\
\hline
zpcfabs\tablenotemark{b} & $N_{\mathrm{H}}$ & $10^{22}$ cm$^{-2}$ & 7.26 $\pm$ 0.10 & 1933/948 & 9401/2 \\
             & Covering Fraction & & 0.984 $\pm$ 0.001 & & \\
             & $z$\tablenotemark{a} & & 0.0426 & & \\
\hline 
mekal & $kT$ & keV & 0.181 $\pm$ 0.004 & 1050/946 & 883/2 \\ 
\hline 
zgauss & Line $E$ & keV & 7.81 $\pm$ 0.06 & 1041/943 & 9/3 \\ 
            & $\sigma$ & keV & $<$0.24 & & \\
            & $z$\tablenotemark{a} & & 0.0426 & & \\
            & EW & eV & -46 $_{-36}^{+29}$ & & \\
\hline 
zgauss & Line $E$ & keV & 6.70 $\pm$ 0.06 & 1028/941 & 13/2 \\
            & $\sigma$\tablenotemark{a} & keV & 0.01 & & \\
            & $z$\tablenotemark{a} & & 0.0426 & & \\       
            & EW & eV & 35 $\pm$ 17 & & 
\enddata
\tablenotetext{a}{Parameters frozen at their stated values.}
\tablenotetext{b}{The covering fraction for the neutral partial covering absorber is purely phenomenological; see \S \ref{sec:broad} for information about a continuum scattering model.}
\end{deluxetable*}

\subsection{Modeling the Iron-K Region}\label{sec:ironK}
To more closely model the iron-K region, we consider the pn and MOS only between $E =$ 2--10 keV, consistent with the methods presented in \citet{xu17} and \citet{braito18}. From Fig. \ref{fig:broad}, it is clear that the residuals are dominated by a complex array of features below $E =$ 2 keV, some of which may be due to a photoionized emitter. Modeling the soft X-ray emission does not impact results from the hard X-ray emission, although the inclusion of the softer energies may detrimentally influence the continuum estimation. Investigating the source of the soft X-ray emission is not our primary objective and is not considered further in this paper. 

\begin{figure*}
\plotone{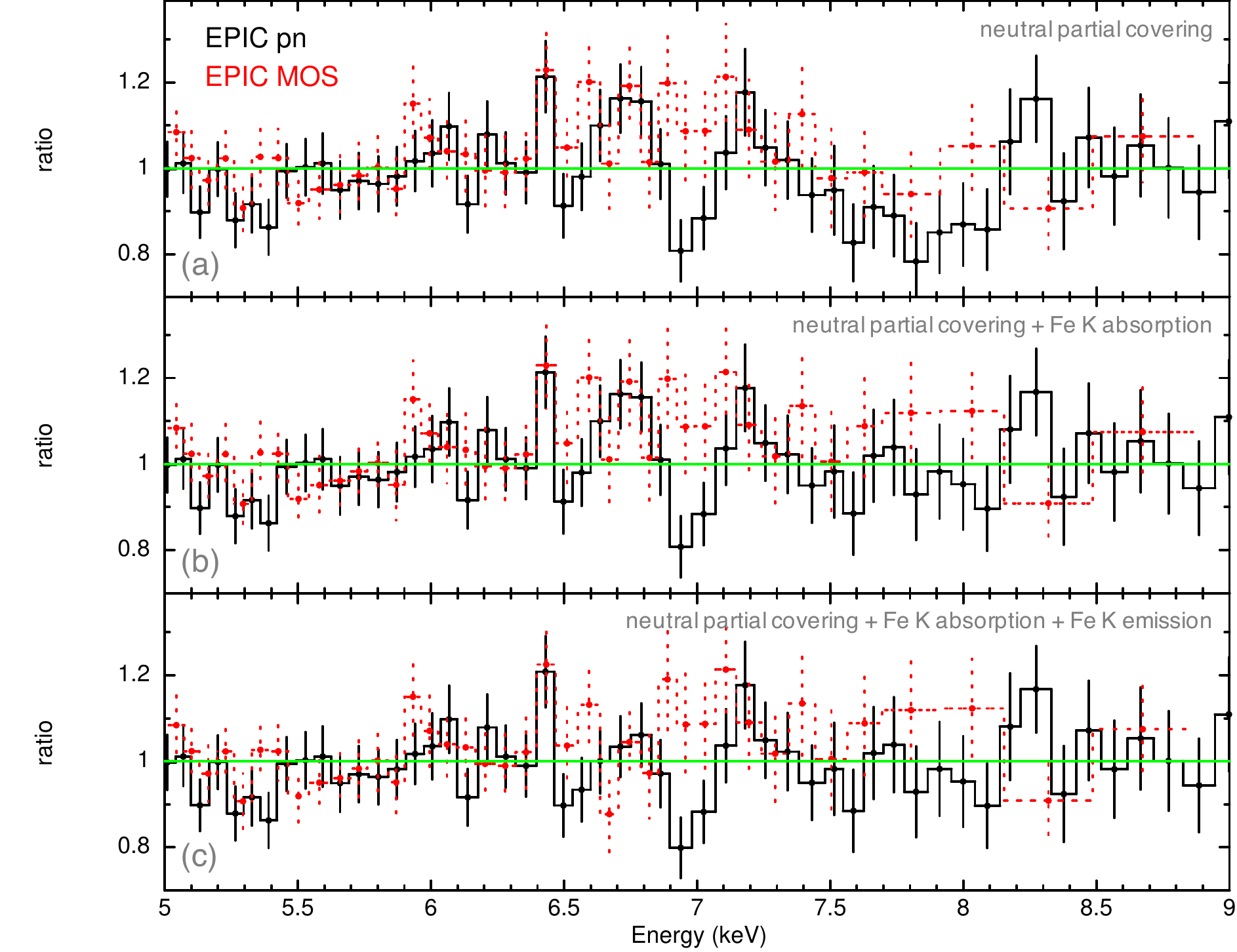}
\caption{Data-to-model ratios for iron-K models ($E$ = 2--10 keV); EPIC pn is plotted in black, the MOS in red. (a) a neutral partial covering absorber model; (b) an Gaussian absorption feature added at $E$ = 7.8 keV which could be indicative of an ultra-fast outflow due to \ion{Fe}{25} He$\alpha$ or \ion{Fe}{26} Ly$\alpha$; (c) a Gaussian emission feature added at $E$ = 6.7 keV. Potential features at $E$ = 6.4 keV, 6.95 keV, and 7.2 keV are not statistically significant.}
\label{fig:210}
\end{figure*}

We begin modeling the iron-K region with a power-law continuum and a neutral partial covering absorber (see Fig. \ref{fig:210}a). We do not include the \texttt{mekal} component, as it does not contribute above $E$ = 2 keV. The column density is $N_{\mathrm{H}} = (7.28\ ^{+0.12}_{-0.27}$) \e{22} cm$^{-2}$ and the photon index $\Gamma$ = 1.94 $\pm\ 0.02$. The covering fraction is 0.98 $\pm\ 0.01$. 

Next, we add a Gaussian absorption feature at $E = 7.81\ \pm\ 0.12$ keV with a width of $\sigma = 143\ ^{+132}_{-98}$ eV and equivalent width of $72\ ^{+42}_{-38}$ eV (see Fig. \ref{fig:210}b). This improves the fit of the model ($\Delta\chi^2$/$\Delta\nu$ = 13/3), and using an F-test, the addition is statistically significant at a level of 99.7\% ($\sim3\sigma$). The column density is $N_{\mathrm{H}} = (6.96\ \pm\ 0.12$) \e{22} cm$^{-2}$ and the photon index $\Gamma$ = 1.88 $\pm\ 0.02$. The covering fraction is 0.984 $^{+0.014}_{-0.011}$ (full covering remains excluded at the 90\% level). 

Finally, we add a second Gaussian emission feature at $E = 6.70\ \pm\ 0.06$ keV with a fixed width of $\sigma$ = 10 eV (see Fig. \ref{fig:210}c). This improves the fit of the model ($\Delta\chi^2$/$\Delta\nu$ = 9/2), and using an F test, the addition is statistically significant at a level of 99.3\%. The addition of a second Gaussian narrows the first Gaussian to $\sigma$ = 117 eV (90\% upper limit $\sigma <$ 257 eV) and equivalent width of 61 $^{+40}_{-38}$  eV. The column density increases slightly to $N_{\mathrm{H}} = (7.03\ ^{+0.12}_{-0.29}$) \e{22} cm$^{-2}$ and the photon index steepens slightly to $\Gamma = 1.91\ \pm\ 0.02$. The covering fraction remains at 0.984 $^{+0.014}_{-0.011}$ with full covering excluded at the 90\% level. The final parameters for this best-fit model are provided in Table \ref{tab:2to10}. 

In order to better assess the significance of the detection of the Gaussian absorption feature at $E =$ 7.81 keV, we run a series of detailed Monte Carlo simulations, according to the procedure described in \citealt{tombesi10}, quantifying the incidence of spurious lines when blindly searching for features between $E =$ 7-10 keV (rest frame). We adopt the best fit model shown in Table \ref{tab:2to10} after removing the Gaussian absorption lines as our baseline model. We simulate a set of 1000 observations with both the EPIC pn and MOS detectors using the same observation times as given in Table \ref{tab:obsinfo} and grouping the spectra to a minimum of 50 counts per bin.

First, we fit the simulated data using our baseline model checking that the best fit values agree within the uncertainties with the input parameters used to generate the data. Then, we look for the probability of detecting an emission or absorption Gaussian feature between $E =$ 7-10 keV due to random fluctuations of the simulated data. To do this, we add a Gaussian component to Model A with a line centroid restricted to fall  between $E =$ 7-10 keV in steps of 0.1 keV. The width of the Gaussian line is free to vary between $\sigma$ = 0-300 eV. The line normalization is left free to vary during the fit between positive and negative values, thus allowing for the presence of emission or absorption features, respectively.

Using the value of $\Delta\chi^2 =$ 13 as the threshold value, we find that 8 out of 1000 ($f =$ 0.008) simulated spectra include spurious lines which improve the fit by a greater or equal amount. We derive the confidence level of the observed absorption line as $p = 1-f = 0.992$, corresponding to 99.2\% or 2.5$\sigma$.

The significant presence of an absorption feature above $E$ = 7 keV could be indicative of an ultra-fast outflow. The strongest highly ionized iron transitions are \ion{Fe}{25} He$\alpha$ ($E$ = 6.697 keV) and He$\beta$ ($E$ = 7.880 keV) and \ion{Fe}{26} Ly$\alpha$ ($E$ = 6.966 keV) and Ly$\beta$ ($E$ = 8.250 keV). For an absorption feature at $E$ = 7.8 keV (rest-frame), only \ion{Fe}{25} He$\alpha$ and \ion{Fe}{26} Ly$\alpha$ would produce an outflow with velocities of $v_{\mathrm{out}}$ = 0.15$c$ and $v_{\mathrm{out}}$ = 0.11$c$, respectively.

We note an apparent narrow absorption feature at $E$ = 7 keV. However, the EPIC pn and MOS data are not consistent at that energy, and any attempt to fit a Gaussian absorption feature is consistent with a width of $\sigma$ = 0 eV. We conclude that this faint absorption feature may be due to random fluctuations. We also note apparent narrow emission features at $E$ = 6.4 and 7.2 keV. These also are not statistically significant with the current data, but they are close to the expected energies for Fe K$\alpha$ and Fe K$\beta$. They will not be considered further here. 

\begin{deluxetable*}{llcccc}[t!]
\tablecaption{Parameters for the best-fit model for 2--10 keV. All errors are given at the 90\% level while limits are given at 90\%.\label{tab:2to10}}
\tablecolumns{6}
\tablewidth{\textwidth}
\tablehead{
\colhead{Component} &
\colhead{Parameter} &
\colhead{Unit} &
\colhead{Model}  & 
\colhead{$\chi^2$/$\nu$} & 
\colhead{$\Delta\chi^2$/$\Delta\nu$} 
}
\startdata
zpowerlw & $\Gamma$ & & 1.91 $\pm$ 0.02 & 4255/772 & \ldots \\
                & $z$\tablenotemark{a} & & 0.0426 & & \\
\hline
zpcfabs\tablenotemark{b} & $N_{\mathrm{H}}$ & $10^{22}$ cm$^{-2}$ & 7.03 $^{+0.12}_{-0.29}$ & 752/770 & 3503/2 \\
             & Covering Fraction & & 0.984 $^{+0.014}_{-0.011}$ & & \\
             & $z$\tablenotemark{a} & & 0.0426 & & \\
\hline 
zgauss & Line $E$ & keV & 7.81 $\pm$ 0.12 & 739/767 & 13/3 \\ 
            & $\sigma$ & keV & $<0.26$ & & \\
            & $z$\tablenotemark{a} & & 0.0426 & & \\
            & EW & eV & -61 $_{-40}^{+38}$ & & \\
\hline 
zgauss & Line $E$ & keV & 6.70 $\pm 0.06$ & 730/765 & 9/2 \\
            & $\sigma$\tablenotemark{a} & keV & 0.01 & & \\
            & $z$ \tablenotemark{a} & & 0.0426 & & \\         
            &  EW & eV & 31 $\pm$ 18 & & \\
\enddata
\tablenotetext{a}{Parameters frozen at their stated values.}
\tablenotetext{b}{The covering fraction for the neutral partial covering absorber is purely phenomenological; see \S \ref{sec:broad} for information about a continuum scattering model.}
\end{deluxetable*}

\subsection{Detailed Photoionization Modeling of the Fe K Absorber} \label{sec:xstar}

We perform a self-consistent photoionization modeling of the Fe K absorber using absorption tables generated with the photoionization code \texttt{XSTAR} \citep{xstar} with standard solar abundances \citep{asplund09}. The output parameters of the \texttt{XSTAR} fit are the column density, ionization parameter, and the observed absorber redshift $z_o$. The ionization parameter is defined as $\xi$ = $L_{\mathrm{ion}}/(nr^2)$ erg s$^{-1}$ cm \citep{tarter69}, where $L_{\mathrm{ion}}$ is the ionizing luminosity from 1--1000 Ry\footnote{1 Ry $\equiv \frac{m_e e^4}{8\varepsilon_0^2h^2} = 13.6$ eV} and and $r, n$ are the distance from the central source and the number density of the gas, respectively. The observed absorber redshift is related to the intrinsic absorber redshift in the source rest frame $z_a$ as (1 + $z_o$) = (1 + $z_a$)(1 + $z_c$), where $z_c$ is the cosmological redshift of the source. The velocity can then be determined using the relativistic Doppler formula, 1 + $z_a$  = [(1 - $\beta$)/(1 + $\beta$)]$^{1/2}$, where $\beta$  = $v/c$.

In order to best fit the observed width of the absorption feature, we consider three absorption tables with turbulent broadening velocities of 1000 km s$^{-1}$, 5000 km s$^{-1}$, and 10,000 km s$^{-1}$. All fits include the neutral partial covering absorber and a Gaussian emission line $E$ = 6.7 keV. The \texttt{XSTAR} absorber well describes the observed absorption feature at $E$ = 7.8 keV without the need for additional Gaussian components.

\begin{figure*}
\plotone{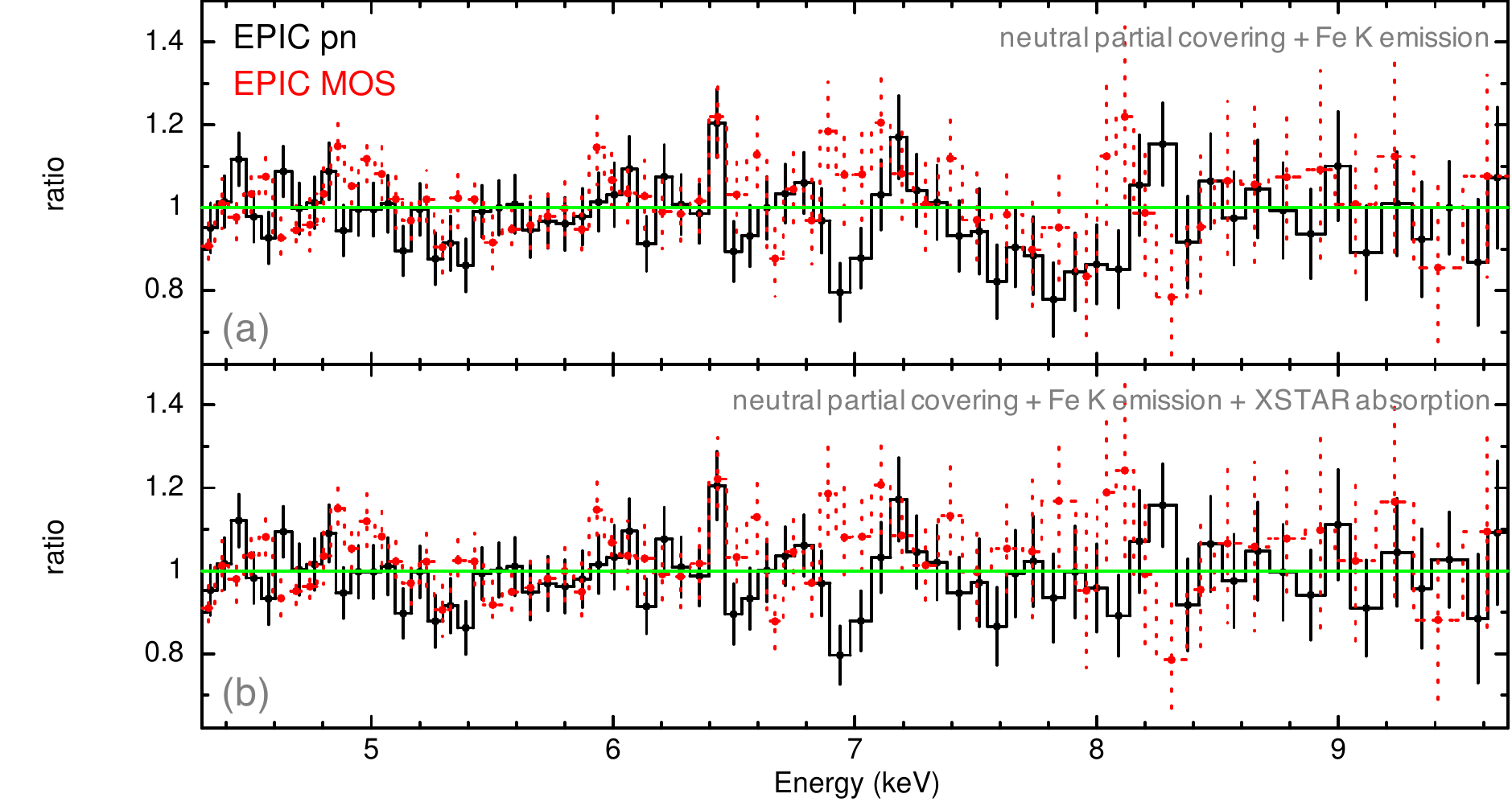}
\caption{Model residuals are presented for two models which both include neutral partial covering and a Gaussian emission line at $E =$ 6.7 keV. The model in the bottom panel includes an \texttt{XSTAR} absorption table which models the absorption feature at $E =$ 7.8 keV.}
\label{fig:5k}
\end{figure*}

\begin{figure}
\epsscale{1.15}
\plotone{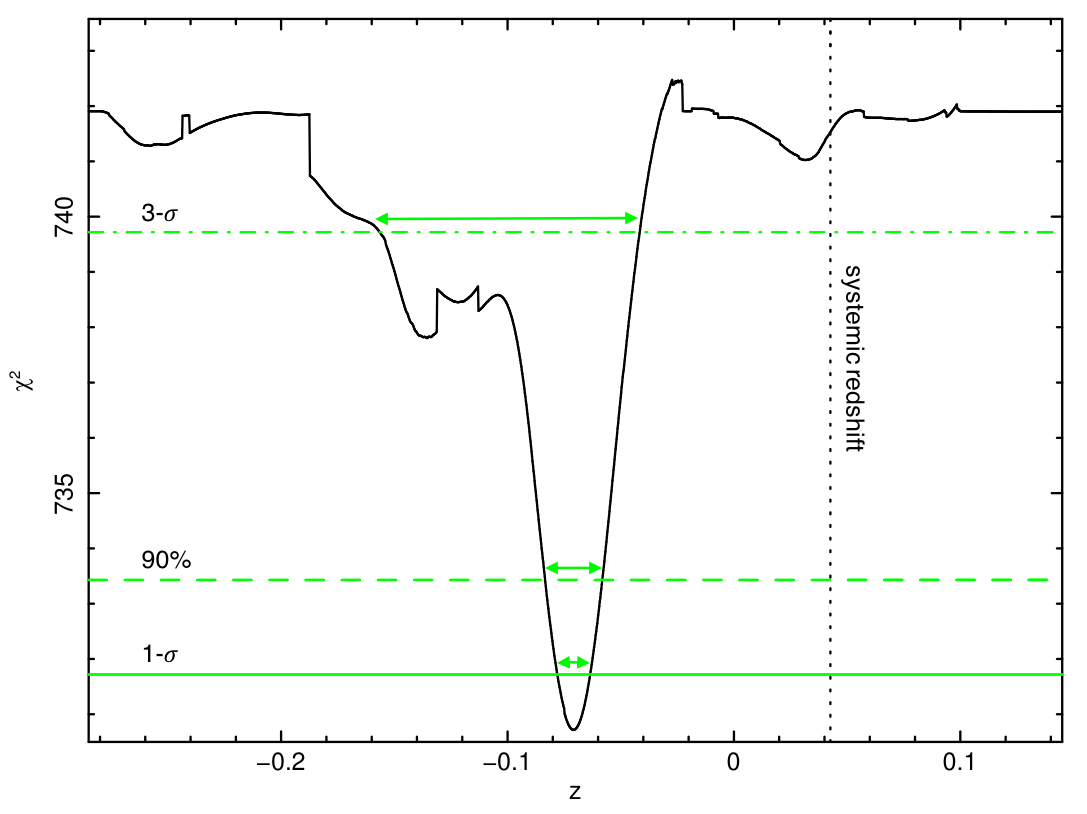}
\caption{Using the XSPEC \texttt{steppar} command, the $\chi^2$ statistic is plotted against the redshift of the \texttt{XSTAR} absorber modeling the Fe K absorption feature at 7.8 keV. The systemic redshift of \obj\ ($z$ = 0.0426) is shown with the vertical dotted line. The solid and dashed horizontal lines indicate the 1-$\sigma$, 90\%, and 3-$\sigma$ confidence ranges for the value of the redshift of the absorber, which is well-constrained at $z$ = -0.07 in the observed frame corresponding to an outflowing velocity of $0.11c$.}
\label{fig:redshift}
\end{figure}

Our best-fit model has a $v_{\mathrm{turb}}$ = 5000 km s$^{-1}$. Model parameters are given in Table \ref{tab:xstar}. Fig. \ref{fig:5k} shows the data-to-model ratios of models with and without the \texttt{XSTAR} component. The redshift of the absorber is well constrained at $z_o= -0.071\ \pm\ 0.012$ (see Fig. \ref{fig:redshift}), which corresponds to an outflowing velocity of $v_{\mathrm{out}}$ = 0.11 $\pm\ 0.01c$. The ionization parameter of $\log\xi = 4.0\ ^{+0.7}_{-0.1}$ erg s$^{-1}$ cm indicates that the absorption feature is due to a mixture of both \ion{Fe}{25} and \ion{Fe}{26} \citep{kallman04}. The covering fraction of the neutral partial covering absorber remains at 0.984 $^{+0.013}_{-0.010}$ with full covering excluded at the 90\% level.

\begin{deluxetable*}{llcccc}[t!]
\tablecaption{Parameters for the best-fitting XSTAR model. All errors are given at the 90\% level while limits are given at 90\%.\label{tab:xstar}}
\tablecolumns{6}
\tablewidth{\textwidth}
\tablehead{
\colhead{Component} &
\colhead{Parameter} &
\colhead{Unit} &
\colhead{Model}  & 
\colhead{$\chi^2$/$\nu$} & 
\colhead{$\Delta\chi^2$/$\Delta\nu$} 
}
\startdata
zpowerlw & $\Gamma$ & & 1.90 $\pm$ 0.02 & 4255/772 & \ldots \\
                & $z$\tablenotemark{a} & & 0.0426 & & \\
\hline
zpcfabs\tablenotemark{b} & $N_{\mathrm{H}}$ & $10^{22}$ cm$^{-2}$ & 6.98 $\pm$ 0.11 & 753/770 & 3502/2\\
             & Covering Fraction & & 0.984 $^{+0.013}_{-0.010}$& & \\
             & $z$\tablenotemark{a} & & 0.0426 & & \\
\hline 
zgauss & Line $E$ & keV & 6.70 $\pm$ 0.06 & 742/768 & 11/2\\ 
            & $\sigma$\tablenotemark{a}  & keV & 0.01 & & \\
            & $z$\tablenotemark{a} & & 0.0426 & & \\
            & EW & & -31 $_{-50}^{+11}$ & &\\
\hline 
XSTAR & $N_{\mathrm{H}}$ & $10^{22}$ cm$^{-2}$ & 26.7 $^{+22.5}_{-12.2}$ & 730/765 & 12/3 \\
            & $\log\xi$ & erg s$^{-1}$ cm  & 4.0 $^{+0.7}_{-0.1}$ & & \\
            & $z$ & & -0.071 $\pm$ 0.012 & & \\
            & $v_{\mathrm{out}}$ & $c$ & 0.11 $\pm$ 0.01 & &          
\enddata
\tablenotetext{a}{Parameters frozen at their stated values.}
\tablenotetext{b}{The covering fraction for the neutral partial covering absorber is purely phenomenological; see \S \ref{sec:broad} for information about a continuum scattering model.}
\end{deluxetable*}

\subsection{Relativistic Reflection Model}\label{sec:rel}

Using previous \xmm\ and \textit{NuSTAR} observations, \citet{xu17} found evidence for relativistic reflection. Although the lack of a clear broad Fe K emission line does not support interpreting the spectrum as dominated by relativistic reflection, we test that possibility in order to compare to the results presented in \citealt{xu17} by using the lamp-post geometry in the \texttt{relxill} code \citep{garcia14, dauser14}. This model (\texttt{relxilllp}) considers a lamp-post geometry in which the compact X-ray emitting source is located on the rotation axis of the black hole at a certain height specified in units of gravitational radii, $R_{\mathrm{g}}$ = $GM_{\mathrm{BH}}/c^2$. The parameters of this model include: (1) $h$, the height of the source in $R_{\mathrm{g}}$, (2) $a$, the dimensionless spin of the black hole, (3) $i$, the inclination with respect to the normal to the accretion disk, (4) $R_{\mathrm{in}}$, the inner radius of the accretion disk, (5) $R_{\mathrm{out}}$, the outer radius of the accretion disk, (6) $z$, the redshift of the system, (7) $\Gamma$, the power law index , (8) $\log\xi$, the ionization parameter of the accretion disk, (9) $A_{\mathrm{Fe}}$, the iron abundance of the accretion disk, (10) $E_{\mathrm{cut}}$, the observed high energy cutoff of the primary spectrum, (11) the reflection fraction (refl\_frac), and (12) a model switch controlling the reflection fraction calculation (fixReflFrac). 

We assume that the inner radius of the accretion disk extends to the ISCO and a typical outer disk radius of $R_{\mathrm{out}}$ = 400 $R_{\mathrm{g}}$. During our analysis, models were not sensitive to the iron abundance, therefore, we fix the iron abundance at solar. We assume an energy cutoff $E_{\mathrm{cut}}$ = 55 keV, the value reported by \citet{xu17} whose spectral analysis extended to $E$ = 30 keV. We set both the reflection fraction and the fixReflFrac switch to 1. Throughout our analysis, $\chi^2$ was minimized by fixing the height of the illuminating source to $h$ = 2 $R_{\mathrm{g}}$, the minimum value permitted by the model. 

A model including a neutral partial covering absorber and \texttt{relxilllp} provides a fit of $\chi^2/\nu$ = 741.5/767. This is not statistically preferred over the models presented in $\S$\ref{sec:ironK} and $\S$\ref{sec:xstar} and is, in fact, worse than our model with an absorbed power law and iron K emission. We still provide full details of this best fit in Table \ref{tab:rel}.

\begin{deluxetable*}{llcccc}[t!]
\tablecaption{Parameters for the best-fitting \texttt{relxilllp} model. All errors are given at the 90\% level while limits are given at 90\%. A full description of model parameters is given in \S\ref{sec:rel}. \label{tab:rel}}
\tablecolumns{6}
\tablewidth{0pt}
\tablehead{
\colhead{Component} &
\colhead{Parameter} &
\colhead{Unit} &
\colhead{Model}  & 
\colhead{$\chi^2$/$\nu$} & 
\colhead{$\Delta\chi^2$/$\Delta\nu$} 
}
\startdata
zpcfabs\tablenotemark{b} & $N_{\mathrm{H}}$& $10^{22}$ cm$^{-2}$ & 7.29 $\pm$ 0.12 & 753/770 & \ldots \\
             & Covering Fraction & & 0.984 $^{+0.013}_{-0.009}$& & \\
             & $z$\tablenotemark{a} & & 0.0426 & & \\
\hline 
relxilllp & $h$\tablenotemark{a} & $R_{\mathrm{g}}$ & $<$16 & 742/767 & \ldots \\
            & $a$ & & 0.62 $^{+0.13}_{-0.25}$ & & \\
            & $i$ & degrees & 49 $\pm$ 4 & & \\
            & $R_{\mathrm{in}}$\tablenotemark{a} & $R_{\mathrm{g}}$ & -1 & & \\
            & $R_{\mathrm{out}}$\tablenotemark{a} & $R_{\mathrm{g}}$ & 400 & & \\ 
            & $z$\tablenotemark{a} & & 0.0426 & & \\
            & $\Gamma$ & & 1.94 $\pm$ 0.03 & & \\
            & $\log\xi$ & erg s$^{-1}$ cm  & 2.3 $\pm$ 0.5 & & \\
            & $A_{\mathrm{Fe}}$\tablenotemark{a} & solar & 1 & & \\
            & $E_{\mathrm{cut}}$\tablenotemark{a} & keV & 55 & & \\ 
            & Reflection Fraction\tablenotemark{a} & & 1 & & \\
            & Fix Reflection Fraction\tablenotemark{a} & & 1 & & \\  
\enddata
\tablenotetext{a}{Parameters frozen at their stated values.}
\tablenotetext{b}{The covering fraction for the neutral partial covering absorber is purely phenomenological; see \S \ref{sec:broad} for information about a continuum scattering model.}
\end{deluxetable*}


\section{\nus\ Spectral Analysis} \label{sec:nuspec}
We perform our spectral analysis using XSPEC v12.10c \citep{arnaud96} using $\chi^2$ statistics. All models take into account Galactic absorption with the \texttt{tbabs} model \citep{wilms00} using a Galactic column density of $N_{\mathrm{H,Gal}} = 1.66\times10^{20}$ cm$^{-2}$ \citep{kalberla05}. All parameters are given in the rest frame of \obj\ ($z$ = 0.0426). All errors and limits are given at the level of 90\% ($\Delta\chi^2 = 2.7$ for one degree of freedom). Statistical calculations were performed using XSPEC \texttt{error} and \texttt{steppar} commands avoiding local minima when searching $\chi^2$ space.

\begin{figure}
\epsscale{1.35}
\plotone{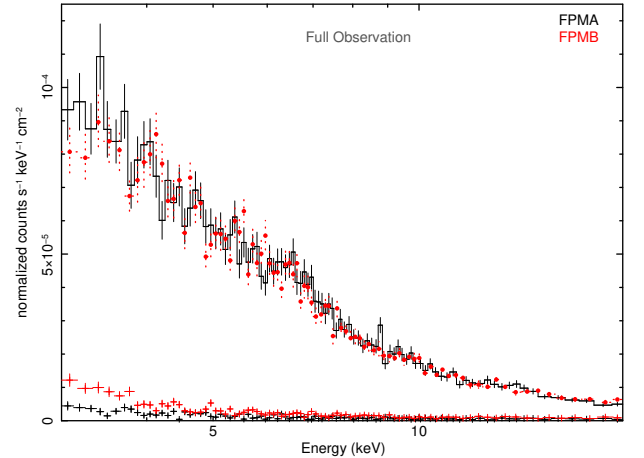}
\caption{Spectra of the full \nus\ observation between $E = 3-20$ keV (rest frame) with the FPMA and B shown in black and red, respectively. The background for each detector is also included.}
\label{fig:nu_spec}
\end{figure}

The \nus\ observation may provide a useful comparison for the results based on the \xmm\ observation. However, we note that while \nus\ may place helpful constraints on the high-energy continuum shape and broad spectral features, the energy resolution of \nus\ is not well suited for the investigation of faint and narrow spectral lines like those found in the Fe K region of the \xmm\ spectra. There is no unusually large flux variability over the course of the full \nus\ observation, however it is important to keep in mind the likely variable nature of UFO absorption features \citep{matzeu16}. Features observed with \xmm\ may or may not be present (or present with the same strength) during the \nus\ exposure that is before and after the \xmm\ observation. The spectra remain signal-dominated until $E$ = 20 keV, but since our goal is to compare with \xmm, we perform our spectral analyses in the mutual energy band from $E$ = 3--10 keV (rest frame). Figure \ref{fig:nu_spec} shows the spectrum and background for the full \nus\ spectrum.

\begin{figure}
\epsscale{1.35}
\plotone{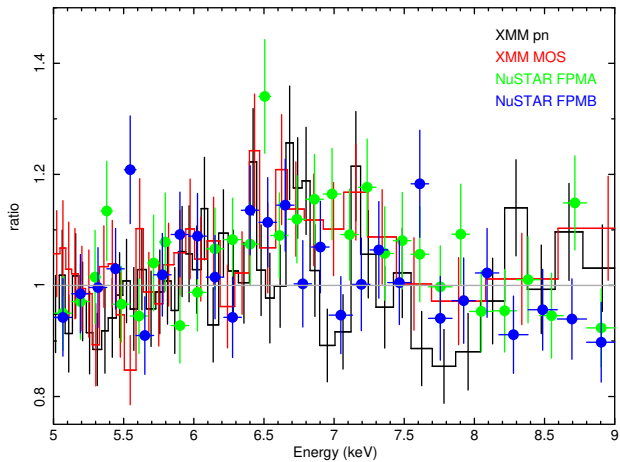}
\caption{Model ratio of the full \nus\ and \xmm\ observations fit with a power-law and neutral partial covering absorber. Additional binning applied for visual purposes.}
\label{fig:xmm_nu_all}
\end{figure}

Due to the short exposure time of the \nus\ observation which is concurrent with \xmm, the signal to noise is not sufficient to detect spectral lines. We therefore focus our analysis on the full \nus\ spectrum. We begin our examination of the full \nus\ spectrum by fitting the data with a power-law continuum and neutral partial covering absorber. Fig. \ref{fig:xmm_nu_all} shows the ratio of this fit along with the \xmm\ observation. We freeze the covering fraction of the neutral partial absorber to 0.984. This corresponds to the best fit value in models of the \xmm\ observation (see \S\ref{sec:spectra}) where the higher sensitivity in the soft energy band (i.e., $E<3$ keV) provides tighter constraints on the covering fraction. Next, we add the two Gaussian features detected in \xmm. Both the central energy values ($E$ = 7.8 keV and 6.70 keV) and the widths ($\sigma$ = 0.12 keV and 0.01 keV) of the Gaussian features were frozen to the values found in \xmm\ because they are could not be constrained with \nus. We do, however, allow the normalization of each Gaussian feature to vary between [-1, 1] keV, thus allowing each Gaussian to be \textit{either} an emission or absorption feature.

This model provides a reasonable fit to the data with a $\Delta \chi^2_{\mathrm{red}}$ = 1.03. We find a steeper photon index of $\Gamma$ = 2.13 $\pm$ 0.09. The data are consistent with either an emission or absorption feature at the energy of $E$ = 7.81 keV, with an equivalent width of 17.1 eV $^{+37.0}_{-40.7}$ eV. Note that at the 90\% level, this is consistent with the \xmm\ detection of an absorption feature, but the feature is not constrained in \nus\ alone. The \nus\ spectrum suggests an emission feature at $E$ = 6.70 keV with an equivalent width of 75 eV $\pm$ 30 eV, also consistent with \xmm\ at the 90\% level.

\section{Discussion} \label{sec:discuss}

\subsection{Accretion Disk Wind} \label{sec:ufo}

In $\S$\ref{sec:spectra}, we report the analysis of the spectrum of \obj\ with a new higher signal-to-noise \xmm\ observation. We find that modeling the Fe K region of the spectrum with a self-consistent photoionization table generated with \texttt{XSTAR} indicates the presence of an outflowing accretion disk wind with a velocity of $v_{\mathrm{out}}$ = 0.11 $\pm\ 0.01c$.

We can estimate the energetics of the wind following the approach described in \citet{tombesi13,tombesi15,tombesi17}. In our study of the energetics, we will use our best-fit model presented in $\S$\ref{sec:xstar} comprised of a neutral partial covering absorber, Gaussian Fe K emission line at $E$ = 6.7 keV, and an \texttt{XSTAR} component modeling the Fe K absorption feature at $E$ = 7.8 keV. 

There are multiple published values for the mass of the central SMBH in \obj \footnote{\citet{dasyra06} derive a dynamical mass estimate of $M_{\mathrm{BH}}$ = 2.95\e{7} $M_\sun$ using CO as a tracer of young stellar velocity dispersions. This method is now understood to systematically underestimate the black hole masses of actively star-forming galaxies like \obj\ because the CO is tracing only the \textit{young} stellar population rather than the older stellar population whose movement is more indicative of the central mass.}. The photometrically derived black hole mass is estimated to be $M_{\mathrm{BH}}$ = 20.8\e{7} $M_\sun$ \citep{veilleux09b}. Using the central velocity dispersions measured from the \ion{Ca}{2} triplet line widths \citep{rothberg13} and the $M_{\mathrm{BH}}-\sigma$ relation \citep{tremaine02}, the mass is estimated to be $M_{\mathrm{BH}}$  = 42\e{7} $M_\sun$ \citep{xu17}. Hereafter, we assume the black hole mass calculated in \citet{veilleux09b} $M_{\mathrm{BH}}$ = 20.8\e{7} $M_\sun$ as a conservative estimate of the black hole mass and thus the X-ray wind energetics.

A lower limit on the radius of the wind can be derived from the radius at which the observed velocity corresponds to the escape velocity, $r_{\mathrm{min}} = 2GM_{\mathrm{BH}}/v^2_{\mathrm{out}} \simeq 5.08\e{15}$ cm. Converting to units of Schwarzchild radii ($R_{\mathrm{S}} = 2GM_{\mathrm{BH}}/c^2$), we obtain a wind launching radius $r \geqslant 83\ R_{\mathrm{S}}$ from the central SMBH. An upper limit on the radius of the wind can be derived from the definition of the ionization parameter ($\xi$) as long as the thickness of the absorber does not exceed its distance to the SMBH, $N_{\mathrm{H}} \simeq n\Delta r < nr$ (e.g. \citealt{crenshaw12}). This assumption is consistent with a disk wind observed close to its launching region. Using the XSPEC \texttt{lumin} command and an unabsorbed power law model, we calculate the ionizing luminosity between 1--1000 Ry to be $L_{\mathrm{ion}}$ = 8.15\e{43} erg s$^{-1}$. Using the column density and ionization parameter from our best-fit model (Table \ref{tab:xstar}) and the definition of the ionization parameter, we find $r_{\mathrm{max}} = L_{\mathrm{ion}} / \xi N_{\mathrm{H}}$ = 3.05\e{16} cm or $r \leqslant 497\ R_{\mathrm{S}}$. 

In calculating the energetics, we will only consider the lower limit on the radius of the UFO. Although the estimate of the upper limit is robust, it is far greater than the true location of the outflow. The mass outflow rate of the wind can be estimated considering the equation $\dot{M}_{\mathrm{out}} = 4 \pi \mu m_{\mathrm{p}} r N_{\mathrm{H}} C_{\mathrm{F}} v_{\mathrm{out}}$ where $\mu$ = 1.4 is the mean atomic mass per proton, $m_{\mathrm{p}}$ is the proton mass, and $C_{\mathrm{F}}$ is the wind covering fraction \citep{crenshaw12}. Assuming spherical symmetry, the solid angle subtended by the wind is $\Omega = 4 \pi C_{\mathrm{F}}$. We conservatively assume $C_{\mathrm{F}} \simeq$ 0.5 estimated from the fraction of sources with detected UFOs and warm absorbers (e.g., \citealt{tombesi10, tombesi13, tombesi14, crenshaw12, gofford13}). Using the range of launching radii calculated above, we find a mass outflow rate of $\dot{M}_{\mathrm{out}} \gtrsim 1.0\ M_\sun$ yr$^{-1}$.

Conservatively assuming that the outflow has reached a terminal velocity, the kinetic (or mechanical) power of the wind can be derived as $\dot{E}_{\mathrm{K}} = \onehalf \dot{M}_{\mathrm{out}} v_{\mathrm{out}}^2 \gtrsim$ 3.6\e{44} erg s$^{-1}$. The momentum rate (or force) of the wind is estimated to be $\dot{P}_{\mathrm{out}} = \dot{M}_{\mathrm{out}} v_{\mathrm{out}} \gtrsim$ 2.2\e{35} dyne. \obj\ has a bolometric luminosity $L_{\mathrm{bol}}$ = 6.47\e{45} erg s$^{-1}$ of which 71\% is attributed to the AGN ($L_{\mathrm{AGN}} = 4.6\e{45}$ erg s$^{-1}$; \citealt{veilleux09b}). Comparing the wind energetics to the AGN luminosity, we find $\dot{E}_{\mathrm{K}} \gtrsim$ 8\% $L_{\mathrm{AGN}}$ and $\dot{P}_{\mathrm{out}} \gtrsim$ 1.4 $L_{\mathrm{AGN}}/c$. These calculated values are in line with those found in studies with larger samples of disk winds in Seyferts and luminous quasars (e.g., \citealt{tombesi12, tombesi15, gofford15, nardini15, fiore17}). The accretion disk wind is consistent with having a momentum rate comparable to the AGN radiation pressure, and the energetics are high enough to influence AGN feedback (e.g. \citealt{dimatteo05,hopkins10,gaspari11}). 

\begin{deluxetable*}{lcccccccc}[t!]
\tablecaption{Location and energetics of the hot ionized disk wind (UFO) and multi-phase galaxy-scale outflows.   Errors are reported when provided in the appropriate references. \label{tab:outflows}}
\tablecolumns{7}
\tablewidth{0pt}
\tablehead{
\colhead{Gas Phase} &
\colhead{$r_{\mathrm{wind}}$ $^{(a)}$}  &
\colhead{$v_{\mathrm{wind}}$ $^{(b)}$} & 
\colhead{$\dot{M}_{\mathrm{wind}}$} &
\colhead{$\dot{P}_{\mathrm{wind}}$} &
\colhead{$\dot{E}_{\mathrm{K,wind}}$} &
\colhead{Reference} \\
 & \colhead{(pc)} & \colhead{(km s$^{-1}$)} & \colhead{($M_\sun$ yr$^{-1})$} & \colhead{($10^{34}$ dyn)} & \colhead{(10$^{42}$ erg s$^{-1})$}  &
}
\startdata
hot ionized & 0.002--0.010 & 33,000 $\pm$ 3000 & 1.0--6.3 & 22--130 & 360--2150 & 1 \\
neutral & 3000 & 560 & 96$^{+12}_{-6}$ & 59$^{+9}_{-4}$ & 38$^{+7}_{-3}$ & 4\\
warm ionized & 3000 & 423 & 2.5$^{+ 0.44}_{-0.69}$ & 0.78$^{+0.14}_{-0.22}$ & 0.21$^{+0.04}_{-0.05}$ & 4\\
molecular (CO) & 189 & 491 & 219 & 68 & 17 & 2\\
molecular (OH)&  &  &  &  &  & \\
\qquad low-velocity & 170 & 200 &120 & 16 & 1.6 & 3\\
\qquad high-velocity & 340 & 550 & 150 & 52 & 14 & 3\\
\qquad total & & & $269^{+19}_{-131}$ & $68^{+14}_{-30}$ & $16^{+4}_{-7}$ & 3\\
\enddata
\tablenotetext{}{References:} 1) This paper; 2) \citealt{fluetsch19}; 3) \citealt{gonzalf17}; 4) \citealt{rupke17}
\tablenotetext{a}{Radius of wind used for calculation of energetics.}
\tablenotetext{b}{Velocities from Reference 1 are the average over all spaxels of the second component central velocity.}
\end{deluxetable*}

\subsection{Connection with Galaxy-scale Outflows} \label{sec:gal}

Galaxy-scale outflows have been observed in \obj\ in neutral, warm ionized, and molecular gas phases \citep{gonzalf17, rupke17, fluetsch19}. Results from the relevant observations are included in Table \ref{tab:outflows}. Energetics derived from the neutral and warm-ionized outflowing gas are based on the ground-based integral field spectroscopy (IFS) of \citet{rupke17}. These observations are limited by the seeing ($\sim$1'') which sets an artificial minimum radius $r\sim$ 400 pc. The adopted radius for the neutral and warm-ionized gas are directly measured from IFS data and are virtually the same. However, a detailed inspection of the neutral and warm-ionized gas phases reveals that they differ in spatial distribution. Note that the warm ionized gas phase is negligible compared to the other phases of the large-scale outflow, so it will not be considered any further in our discussion.

The energetics for the molecular outflows are derived using OH and CO as tracers for H$_2$. OH absorption features are detected against the unresolved continuum emission in \textit{Herschel} far-infrared spectra \citep{gonzalf17}. The dimensions and energetics of the OH outflow are derived by carefully comparing the velocity profiles of four ground-state and radiatively excited transitions of OH and the predictions from spherically symmetric radiative transfer models. OH molecular tracers are sensitive to the dense molecular gas in the nucleus, so this gas component does not extend much beyond $r\sim$ 500 pc. The CO energetics are derived from millimeter wave interferometry of spatially resolved CO emission lines, and the adopted radius is directly measured from these data \citep{fluetsch19}. As seen in Table \ref{tab:outflows}, there is good agreement between the different tracers. To simplify our discussion of the energetics, we take the average of the momentum rates for the neutral and two molecular outflows ($\dot{P}_{\mathrm{out,av}}$ = 65\e{34} dyn) since these phases likely provide measurements of the same outflow at different epochs (i.e. distances from the center). We similarly take the average of the outflow velocity for the neutral, molecular CO, and high-velocity molecular OH outflows ($v_{\mathrm{out,av}}$ = 534 km s$^{-1}$).

To compare the energetics of the X-ray outflow with the galaxy-scale outflow, we consider two different ways to drive a galaxy-scale outflow. In the case of a momentum-driven outflow, we expect $\dot{P}_{\mathrm{outer}} \simeq \dot{P}_{\mathrm{inner}}$ where ``outer" refers to the galaxy-scale outflow and ``inner" refers to the inner X-ray wind \citep{zubovas12, faucher12}. In \S\ref{sec:ufo}, we derive $\dot{P}_{\mathrm{inner}} \gtrsim$ 22\e{34} dyn, while the momentum rate for the galactic scale outflows are consistently measured as $\dot{P}_{\mathrm{outer}} \sim$ 65\e{34} dyn (Table \ref{tab:outflows}). This gives $\dot{P}_{\mathrm{outer}}/\dot{P}_{\mathrm{inner}} \sim$ 3, however, given the large uncertainties in the momentum rate estimates, our data are not inconsistent with a momentum-driving scenario. 

For an energy-driven outflow, conservation of energy gives $\frac{1}{2}\dot{M}_{\mathrm{inner}}v_{\mathrm{inner}}^2 = \frac{1}{2}f\dot{M}_{\mathrm{outer}}v_{\mathrm{outer}}^2$ where ``outer" refers to the galaxy-scale outflow and ``inner" refers to the inner X-ray wind. The efficiency factor, $f$, is limited to [0,1] where $f=0$ and $f=1$ are two extremes indicating either full dissipation or conservation of kinetic power within the outflow, respectively. Using the expression for the momentum rate, this can be rewritten as $\dot{P}_{\mathrm{inner}}v_{\mathrm{inner}} = f\dot{P}_{\mathrm{outer}}v_{\mathrm{outer}}$. Thus, the expected momentum rate for the large-scale outflow in an energy-driven outflow is given as $\dot{P}_{\mathrm{outer}} = f(v_{\mathrm{inner}}/v_{\mathrm{outer}})\dot{P}_{\mathrm{inner}}$. The efficiency factor can be interpreted as the ratio between the covering fractions of the inner and outer outflows or the fraction of the kinetic energy of the inner X-ray wind that goes into bulk motion of the swept-up molecular material. 

Using average values for the large-scale galactic outflows along with the lower limit of the momentum rate for the UFO, we obtain $f$ = 0.05. This low efficiency value could be the result of a highly clumpy interstellar medium or if the covering fraction of the large-scale outflow is low \citep{wagner12,wagner13,hopkins16}. We note that the ratio of the momentum rate of the molecular outflow to the momentum rate of UFO ($\dot{P}_{\mathrm{mol}}/\dot{P}_{\mathrm{UFO}} \sim$ 0.5-3) is approximately of order unity within the errors. \citet{richings18} find that $\dot{P}_{\mathrm{mol}}/\dot{P}_{\mathrm{UFO}}$ of order unity could still be attributed to an energy-driven outflow where the thermalized mechanical energy is mostly lost through efficient cooling due to in-situ formation of molecular gas within the outflow.

Additionally, we note that a purely IR radiation driven molecular outflow (as opposed to mechanical acceleration; see e.g. \citealt{kingpounds15}) is not preferred, but not strictly ruled out. In such a scenario, the momentum of the molecular outflow is given by $\dot{P}_{\mathrm{mol}}\sim (1+\eta\tau_{\mathrm{IR}})(L_{\mathrm{IR}}/c)$ where theoretically $\eta \sim$ 0.5--0.9 \citep{zhang17, ishibashi18} and $\tau_{\mathrm{IR}}$ is the optical depth in the infrared. For \obj, $L_{\mathrm{IR}}$ = 1.38\e{12} $L_{\sun}$ \citep{gonzalf17} which implies $\tau_{\mathrm{IR}} \sim$ 3--5, and thus requires significant IR trapping. 

\begin{figure*}
\plotone{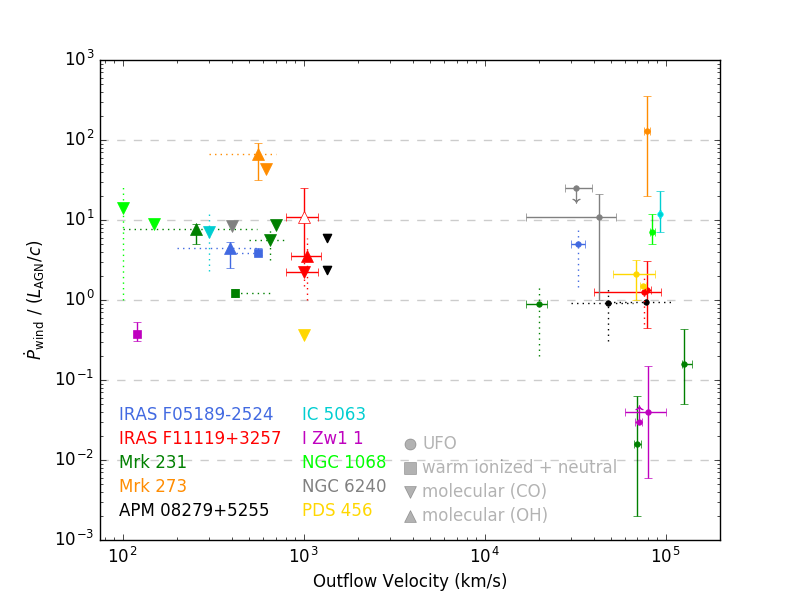}
\caption{The momentum rate ($\dot{P}_{\mathrm{wind}}$) normalized by the momentum of the radiation ($L_{\mathrm{AGN}}/c$) is plotted against the wind outflow velocity for ten objects with observed ultra-fast outflows and large-scale galactic outflows with good constraints on their spatial scales. Solid error bars indicate that upper and lower errors were calculated whereas dotted error bars indicate that only a range of values was provided. Arrows indicate limits. UFO measurements are plotted as circles, warm ionized and neutral gas as squares, the molecular (CO) as downward triangles, and the molecular (OH) as upward triangles. For molecular measurements, filled symbols indicate a time-averaged momentum rate whereas an open symbol is an ``instantaneous" or local momentum rate. See Table \ref{tab:energetics} in Appendix \ref{sec:energetics} for more details and references for each specific object.}
\label{fig:energetics}
\end{figure*}

Finally, we consider \obj\ in the context of nine other sources which have observed UFOs and large-scale galactic outflows with good constraints on their spatial scales. Fig. \ref{fig:energetics} shows the momentum rate against the velocity of the outflow while Appendix \ref{sec:energetics} includes detailed information and references for each object. It is clear that some objects reside in the momentum-driven regime while others are more consistent with the energy-driven scenario suggesting that there is a range of efficiency factors ($f\sim$ 0.001--0.5) that likely depend on specific physical conditions in each object.

\section{Conclusions} \label{sec:con}

We present new \xmm\ and \nus\ observations of the galaxy merger \obj\, a ULIRG and optical Seyfert 2. Testing multiple spectral models yields a best-fit model consisting of a highly ionized absorber with either an absorbed power law and neutral partial covering absorber or a neutral absorber and scattered emission. We find evidence for a blueshifted Fe K absorption feature at $E$ = 7.8 keV (rest-frame) which implies an ultra-fast outflow with $v_{\mathrm{out}} = 0.11 \pm\ 0.01c$. 

We calculate that the UFO has a mass outflow rate of $\dot{M}_{\mathrm{out}} \gtrsim$ 1.0 $M_\sun$ yr$^{-1}$, a kinetic power of $\dot{E}_{\mathrm{K}} \gtrsim$ 3.6\e{44} erg s$^{-1}$ (8\% $L_{\mathrm{AGN}}$), and a momentum rate (or force) of $\dot{P}_{\mathrm{out}} \gtrsim$ 22\e{34} dyne (1.4 $L_{\mathrm{AGN}}/c$). Observed large-scale galactic outflows in \obj\ have an average momentum rate of $\dot{P}$ = 68\e{34} dyne, yielding $\dot{P}_{\mathrm{inner}}/\dot{P}_{\mathrm{outer}} \sim 3$. Given the large uncertainties in the momentum rate estimates, $\dot{P}_{\mathrm{inner}}/\dot{P}_{\mathrm{outer}}$ is not inconsistent with unity, or a momentum-driven scenario. In the energy-driven outflow scenario, the fraction $f$ of the kinetic energy in the inner X-ray wind that goes into bulk motion of the large-scale outflow is $f \sim 0.05$. Such a low efficiency could be attributed to a highly clumpy interstellar medium or if the covering fraction of the large-scale outflow is low or if the hot gas has efficiently cooled leading to an in-situ formation of the molecular outflow.

We compare the outflow in \obj\ to nine other objects with observed UFOs and large-scale galactic outflows with solid constraints on the outflow energetics. We find that there is a range of efficiency factors ($f\sim$ 0.001--0.5) and driving mechanisms that likely depend on specific physical conditions in each object. While this remains a small sample, it is an important step towards building a comprehensive sample which can be used to further probe the complex relationships of AGN and galaxy co-evolution.

\acknowledgments
R.N.S. and S.V. acknowledge support from NASA, grant number 15-NUSTAR215-0021. F.T. acknowledges support by the Programma per Giovani Ricercatori - anno 2014 `Rita Levi Montalcini'. S.V. acknowledges support from a Raymond and Beverley Sackler Distinguished Visitor Fellowship and thanks the host institute, the Institute of Astronomy, where this work was concluded. S.V. also acknowledges support by the Science and Technology Facilities Council (STFC) and by the Kavli Institute for Cosmology, Cambridge. A.L. acknowledges financial support from the Italian Space Agency (ASI) under the contract ASI-INAF  n.2017-14-H.0.

\facilities{\xmm\ (EPIC)}

\clearpage

\appendix

\section{Outflow Energetics of the AGN Sample}\label{sec:energetics}

\startlongtable
\begin{deluxetable}{l l l l l}
\tablecaption{Outflow velocities and momentum rates for ten objects with observed ultra-fast outflows and large-scale galactic outflows with good constraints on their spatial scales. Errors are presented when published by their respective authors. For simplicity, objects with both warm ionized and neutral observed outflows were plotted as a single data point in Fig. \ref{fig:energetics}. \label{tab:energetics}}
\tablecolumns{4}
\tablewidth{0pt}
\tablehead{
\colhead{Object} &
\colhead{Gas Phase} &
\colhead{$v_{\mathrm{wind}}$} &
\colhead{$\dot{P}_{\mathrm{wind}}\ ^a$} &
\colhead{Ref.}\\
 & & \colhead{(km s$^{-1}$)} & \colhead{($L_{\mathrm{AGN}}/c$)} &
}
\startdata
 \obj & hot ionized & 33,000 $\pm$ 3,000 & 1.44--8.48 & 1 \\
 & neutral & 560 & 3.85$^{+0.59}_{-0.26}$ & 12 \\
 & warm ionized & 423 & 0.05 $\pm$ 0.01 & 12 \\
 & molecular (CO) & 491 & 4.44 & 6 \\
 & molecular (OH) & 200--550 & 4.44$^{+0.91}_{-1.96}$ & 8\\
IRAS F11119+3257 & hot ionized & 76,500 $\pm$ 3,300 & 1.30$^{+1.70}_{-0.90}$ & 13\\
 & hot ionized & 76,000$^{+18,000}_{-35,000}$ & 0.5--2 & 14\\
 & molecular (CO)& 1000 $\pm$ 200 & 1.5--3.0 & 13\\
 & molecular (OH) &1000 $\pm$ 200 & 11$^{+14.1}_{-7.5}$ & 15 \\
 & molecular (OH) &1000 $\pm$ 200 & 1.0--6.0 & 15\\
Mrk 231 & hot ionized & 20,000$^{+2,000}_{-3,000}$ & 0.2--1.6 & 4$^b$ \\
 & hot ionized & 127,000$^{+13,000}_{-4,000}$ & 0.16$^{+0.27}_{-0.11}$ & 10 \\
 & hot ionized & 70,000 $\pm$ 3,000 & 0.016$^{+0.048}_{-0.014}$ & 10 \\
 & neutral & 416 & 1.23$^{+0.15}_{-0.08}$ & 12 \\
 & warm ionized & 672 & 0.008 $\pm$ 0.001 & 12 \\
 & molecular (CO) & 500--800 & 3.2--8.0 & 4$^b$ \\
 & molecular (CO) & 700 & 8.7 & 3$^b$ \\
 & molecular (OH) & 100--550 & 7.74$^{+2.68}_{-1.05}$ & 8 \\
 Mrk 273 & hot ionized & 79,000 $\pm$ 3,000 & 130$^{+220}_{-110}$& 10 \\
 & molecular (CO) & 620 & 43 & 3$^b$ \\
 & molecular (OH) & 300--700 & 67$^{+25}_{-35}$ & 8 \\
APM 08279+5255 & hot ionized & 48,000--108,000 & 0.95  & 5 \\
 & hot ionized & 30,000--66,000 & 0.3--1.5 & 5 \\
 & molecular (CO) & 1340 & 2.37 & 5 \\
 & molecular (CO) & 1340 & 5.97 & 5 \\
IC 5063 & hot ionized & 93,000$^{+1,300}_{-1,400}$ & 12$^{+11}_{-5}$ & 10\\
 & molecular (CO) & 300 & 2.3--12.0 & 3$^b$ \\
I Zw 1 & hot ionized & 80,000 $\pm$ 20,000 & 0.04$^{+0.11}_{-0.03}$ & 10 \\
& hot ionized & 71,000 $\pm$ 3,000 & $>$0.03 & 10 \\
& neutral & 120 & 0.37$^{+0.15}_{-0.07}$ & 12 \\
NGC 1068 & hot ionized & 84,000$^{+3,000}_{-2,000}$ & 7$^{+5}_{-2}$ & 10 \\
& molecular (CO) & 100 & 1--27 & 7 \\
& molecular (CO) & 150 & 9 & 3$^b$ \\
NGC 6240 & hot ionized & 43,000$^{+10,000}_{-26,000}$ & 11 $\pm$ 10 & 10 \\
& hot ionized & 32,000$^{+7,000}_{-4,000}$ & $<$25 & 10 \\
& molecular (CO) & 400 & 8 & 3$^b$\\
PDS 456 & hot ionized & 69,000 $\pm$ 18,000 & 2.1 $\pm$ 1.1 & 9\\
& hot ionized & 75,000 $\pm$ 3,000 & 1.5 & 11 \\
& molecular (CO) & 1000 & 0.36 & 2  \\ 
\enddata
\tablenotetext{}{References: 1) This paper; 2) \citealt{bischetti19}; 3) \citealt{cicone14}; 4) \citealt{feruglio15}; 5) \citealt{feruglio17}; 6) \citealt{fluetsch18}; 7) \citealt{garcb14}; 8) \citealt{gonzalf17}; 9) \citealt{luminari18}; 10) \citealt{mizumoto19}; 11) \citealt{nardini15}; 12) \citealt{rupke17}; 13) \citealt{tombesi15}; 14) \citealt{tombesi17}; 15) \citealt{veilleux17}}
\tablenotetext{a}{$\dot{P}_{\mathrm{wind}} = \dot{M}_{\mathrm{wind}}v_{\mathrm{wind}}$; $\dot{E}_{\mathrm{wind}} = \frac{1}{2}\dot{M}_{\mathrm{wind}}v_{\mathrm{wind}}^2$}
\tablenotetext{b}{CO-based molecular outflow momentum rates from these references were divided by a factor of 3 so that they are on the same scale as the other measurements.}
\end{deluxetable}



\end{document}